# Quantum electron liquid and its possible phase transition

Sunghun Kim[1,†], Joonho Bang[2,†], Chan-young Lim[1,†], Seung Yong Lee[2], Jounghoon Hyun[1], Gyubin Lee[1], Yeonghoon Lee[1], Jonathan D. Denlinger[3], Soonsang Huh[4,5], Changyoung Kim[4,5], Sang Yong Song[6], Jungpil Seo[6], Dinesh Thapa[7], Seong-Gon Kim[7], Young Hee Lee[2,8], Yeongkwan Kim[1,*], and Sung Wng Kim[2,8,*]

[1]Department of Physics, Korea Advanced Institute of Science and Technology, Daejeon 34141, Korea.

[2]Department of Energy Science, Sungkyunkwan University, Suwon 16419, Korea.

[3]Advanced Light Source, Lawrence Berkeley National Laboratory, Berkeley, California 94720, USA.

[4]Center for Correlated Electron Systems, Institute for Basic Science, Seoul 08826, Korea.

[5]Department of Physics and Astronomy, Seoul National University, Seoul 08826, Korea.

[6]Department of Emerging Materials Science, DGIST, Daegu 42988, Korea.

[7]Department of Physics & Astronomy and Center for Computational Sciences, Mississippi State University, Mississippi States, Mississippi 39792, USA.

[8]Center for Integrated Nanostructure Physics, Institute for Basic Science, Sungkyunkwan University, Suwon 16419, Korea.

†These authors contributed equally to this work.

*E-mail: yeongkwan@kaist.ac.kr, kimsungwng@skku.edu

**Purely quantum electron systems exhibit intriguing correlated electronic phases by virtue of quantum fluctuations in addition to electron-electron interactions. To realize such quantum electron systems, a key ingredient is dense electrons decoupled from other degrees of freedom. Here, we report the discovery of a pure quantum electron liquid, which spreads up to ~ 3 Å in the vacuum on the surface of electride crystal. An extremely high electron density and its weak hybridisation with buried atomic orbitals evidence the quantum and pure nature of electrons, that exhibit a polarized liquid phase as demonstrated by our spin-dependent measurement. Further, upon enhancing the electron correlation strength, the dynamics of quantum electrons changes to that of non-Fermi liquid along with an anomalous band deformation, suggestive of a transition to a hexatic liquid crystal phase. Our findings cultivate the frontier of quantum electron systems, and serve as a platform for exploring correlated electronic phases in a pure fashion.**

Electron phases, ranging from gas to liquid and solid, are foundational in physics, chemistry, and materials science. Understanding the characteristics of each electron phase, both from theoretical and experimental viewpoints in a complete range of densities and temperatures, is essential and has been a reference for diverse properties in the actual solid systems[1–3]. In particular, the study of possible phases that can be formed only by electrons spread in 2D space has been a central subject, benefitting from its conceptual simplicity as well as from its expandability toward emergent phenomena such as the Wigner crystal, a solid consisting only of electrons[4].

Non-degenerate 2D electrons formed on the surface of liquid helium (LHe)[5–14] is one platform to study such pure electron phases. The Wigner crystal was first evidenced by examining the transport property of electrons across the phase transition[8,9] and further the possible signature of the hexatic phase, an intermediate liquid crystalline phase, was also captured[12,13]. However, the density of electrons on the surface of LHe only reaches the electrodynamic limit of $< 10^{10}$ cm$^{-2}$, which is restricted in a classical regime[7–9,14]. Yet, degenerate electrons with high density in the quantum regime, which would be far more intriguing, have remained unvisited experimentally. In the degenerate quantum regime, the activated quantum fluctuation dresses the electrons, leading to exotic correlated phases in the vicinity of the Wigner crystal phase, which are not expected in the classical regime. In addition to the various liquid crystalline, stripe, and bubble phases[15], even the superconducting and supersolid phases – the prime problems of condensed matter physics – are possible intermediate phases in the quantum regime[16] which remain to be substantiated experimentally.

To realize pure 2D quantum electron systems, the prerequisite is decoupling the high-density 2D confined electrons from other degrees of freedom. Decoupling is possible when electrons are captured on the positively charged surface of solid matter. Such quantum electrons can be found at the surface of 2D electride crystals[17–19], which consist of positively charged cationic layers and counteranionic electrons located in the interlayer space. Interstitial anionic electrons (IAEs) are localized at the interstitial space between cationic layers and are not bound to the atomic orbitals of neighbouring cations. When the positively charged cationic layer is terminated at the surface of 2D electrides, the IAEs are inevitably detained on the terminated cationic layers to maintain charge neutrality as the distinct surface state from IAEs. It is notable from the geometrical aspect that the theoretical maximum density of the electrons on the surface can be as high as $\sim 9.5 \times 10^{14}$ cm$^{-2}$, which is expected from the concentration of IAEs in a single interlayer space of the 2D electrides with a bulk electron density of $1.4 \times 10^{22}$ $\sim 2.9 \times 10^{22}$ cm$^{-3}$ (ref. 17–19). Through the experiments, the uncharted high-density pure 2D electron system with a concentration of $\sim 2.0 \times 10^{14}$ cm$^{-2}$ is confirmed to be realized on the surface of 2D [Gd$_2$C]$^{2+}\cdot$2e$^-$ electride at high temperatures around 10 K, extending the phase diagram of pure 2D electrons to the quantum regime (Fig. 1a).

What is more surprising is that the pure 2D quantum electrons guarantee a weak hybridisation with the atomic orbitals of the outermost Gd elements of 2D [Gd$_2$C]$^{2+}\cdot$2e$^-$ electride. Density functional theory (DFT) calculations clearly show two distinct electronic states of localized IAEs at the interlayer and surface electrons on the terminated cationic layer, distinguished by the electron localisation function (ELF). Figure 1b–d shows the ELFs of the 2D [Gd$_2$C]$^{2+}\cdot$2e$^-$ electride with strongly localized IAEs and finitely delocalized surface electrons. While the IAEs have local maxima between cationic layers, the ELF on the surface spreads up to $\sim 3$ Å from the outermost Gd atoms, revealing the existence of loosely confined 2D electrons. In experiments using the 2D [Gd$_2$C]$^{2+}\cdot$2e$^-$ electride[19] (Extended Data Fig. 1), we verify the formation of pure 2D quantum electrons by exposing the localized IAEs on the surface of the outermost cationic layer via in situ cleaving at 10 K under an ultrahigh vacuum (UHV) better than $4 \times 10^{-11}$ Torr. Angle-resolved photoemission spectroscopy (ARPES) clearly visualizes a parabolic energy band near the $\Gamma$-point dispersing within the low binding energy region of ~ 250 meV from the Fermi energy ($E_F$) and a V-shaped band at higher energies (Fig. 2a). While the V-shaped band and other complex bands come from the IAEs and the cationic layers, respectively[19], the parabolic band comes from the surface electrons above the cationic layer (see Supplementary Figure 1) and has a 2D nature supported by a negligible dispersion along the $k_z$-axis and isotropic nature by a circular Fermi surface topology in the $k_x$–$k_y$ plane (Extended Data Figs. 2 and 3).

Most importantly, the pure nature of 2D surface electrons, i.e., being free from the orbital network, is demonstrated by the negligible contribution of the Gd atomic orbital to the parabolic band, which is identified by DFT calculations and further confirmed by experimental observations. Figure 2b–d represents the fat band analysis that gives a clear feature for the band characters of the surface electrons and the trapped IAEs inside the bulk. The bands of the IAEs have contributions of Gd atomic orbitals, mostly from the 5$d$ orbital (Fig. 2b) and very weakly from the 6$s$ and 4$f$ orbitals (Fig. 2c,d). Different from the IAE bands, there is almost no Gd orbital contribution on the parabolic band, indicating negligible overlap between surface electrons and underneath Gd orbitals, contradicting a conventional surface state in which electrons are bound around the atomic orbitals. From the resonance behaviour in photoemission intensity for each band, distinct degrees of Gd orbital contribution are identified. Band dispersions observed at photon energies of 144 eV and 148 eV (Fig. 2e,f), which correspond to off- and on-resonant conditions of the Gd 4$d$ core level, respectively, show that the IAE band exhibits resonance behaviour; the dramatically increased intensity under the on-resonant condition ($I_{IAE}$, brown in Fig. 2g) is caused by the finite overlap with Gd orbital. Meanwhile, the parabolic band of surface electrons gives almost constant intensity ($I_{Surface}$, blue in Fig. 2g), which is more apparent in the intensity ratio of $I_{Surface}/I_{IAE}$ (black in Fig. 2g), confirming a very weak hybridisation between their wave functions and the outermost Gd orbitals, a pure nature of 2D surface electrons.

Having evidence that these surface electrons are successfully isolated from other degrees of freedom of the electride solid, the next questions to be addressed are whether the pure electron system is within the desired quantum regime and which quantum phase evolves by electron-electron interactions. Fitting the parabolic band gives an effective mass ($m^*$) of ~ 2.1 $m_e$ and extremely high density ($n$) of ~ $2.0 \times 10^{14}$ cm$^{-2}$ (Fig. 3a,b), in contrast to other 2D electron gases bound to the atomic nuclei with a small $m^*$ value in the range of 0.5–1.4 $m_e$ (ref. 20–22) and to the surface electrons on the LHe with a low density of $< 10^{10}$ cm$^{-2}$ in the classical regime[7–9,14]. The extracted scattering rate of the surface electrons obtained by measuring the peak width in the spectra reveals the liquid nature with a clear quadratic energy dependence, which corresponds to the behaviour of a Fermi liquid[23,24] (Fig. 3c). The high density that surpasses the critical boundary (the solid red line in Fig. 1a) verifies that the electron liquid is in the quantum regime. Furthermore, our quantum electron liquid is spin-polarized, as verified by the spin-resolved measurements that support the spin-polarized band of the surface electrons, which is also revealed by DFT calculations (Fig. 3d). Four sets of spin-dependent spectra are obtained for the surface electrons and the trapped IAEs along the out-of-plane and in-plane directions (Fig. 3f–i). It is evident that spin polarisation of the quantum electron liquid occurs along the out-of-plane direction. Meanwhile, the IAEs have spin polarisation along the in-plane direction, corresponding to the magnetic easy-axis of ferromagnetic bulk [Gd$_2$C]$^{2+}$·2e$^-$ electride, which is induced by the exchange interaction of magnetic quasi-atomic IAEs with Gd atoms of cationic layers[19]. We note that the different direction of spin polarisation between the quantum electron liquid and trapped IAEs implies that the polarized nature of surface electrons is not relevant to the magnetic moment of the underlying bulk but is likely induced by the intrinsic electron-electron interaction, as it is one of the predicted ground states of pure 2D quantum electron phases[25–27]. Therefore, in short, an unprecedented pure 2D quantum electron liquid is demonstrated, which substantiates that the polarized Fermi liquid can emerge between the paramagnetic Fermi liquid and crystalline electron phases in the quantum regime[25–27] (Fig. 1a).

Realising the pure 2D quantum electron liquid assures that the exotic phases expected in the quantum regime with lower electron density and thus stronger electron correlation, are ready to be explored. As a trigger for enhancing the electron correlation, potassium (K) atoms were deposited to the system which usually change the electron density at the surface. Due to the ultralow work function of the electride system, particularly for the loosely bound surface state to the topmost atomic layer, K deposition would decrease the electron density of the quantum liquid. Indeed, the signatures of enhanced electron correlation strength were captured as discussed below, suggesting the reduction of electron density by K deposition. The complete disappearance of the surface electron state below $E_F$ in the measured bands at higher K coverages (Extended Data Fig. 4), as well as the consumption of the surface electrons by deposited K atoms verified by DFT calculations with different coverages of K overlayer (Supplementary Figure 2) also demonstrate the reduction in density. Upon progressive K deposition, we observed an anomalous band deformation of the quantum electrons from the initial parabolic dispersion (Fig. 4a) with an isotropic circular Fermi surface topology (Fig. 4b) to a W-shaped dispersion (Fig. 4a) with an anisotropic hexagonal topology (Fig. 4c and Extended Data Fig. 5).

Next, we discuss the origin of the anomalous W-shaped band dispersion compared to the conventional band renormalisations. The deformed hexagonal topology of the quantum electrons indicates the breaking of continuous rotational symmetry. Meanwhile, when the spatial ordering of deposited K atoms occurs on the surface, band renormalisation, such as band folding, can take place, inducing the deformation of the parabolic band. However, we rule out the band folding effect on the W-shaped band dispersion because the replica bands do not appear in the higher orders of the reduced Brillouin zone (BZ), which should occur if the deposited K atoms order and the quantum electrons are strongly affected by the K ordering (Extended Data Fig. 6). The intact V-shaped IAE band also supports that the ordering of deposited K atoms is unlikely.

Another possibility is that the K deposition can initiate the effect of the underlying lattice, which inevitably induces the band deformation of the quantum electrons into the V-shape around zero momentum, mimicking the IAE band (Extended Data Fig. 7). However, we observed a W-shaped band dispersion around zero momentum, precluding the lattice potential effect of the underlying lattice. In addition, the nontrivial angular warping of the band that cannot be accounted for by the rotational symmetry of the underlying lattice provides further evidence against the lattice potential effect (Extended Data Fig. 5). Indeed, the absence of resonance behaviour of quantum electrons even after K deposition (Extended Data Fig. 8) strongly supports the fact that the surface electrons are persistently in the loosely bound state and thus that the lattice potential effect on the band deformation is negligible. Therefore, the origin of the W-shaped band is not attributed to the spatially modulating potentials of deposited K atoms or the terminated cationic layer. It should be noted that the observed deformation of the band dispersion, predominantly at zero momentum, occurs hardly from the change of translational symmetry.

Instead, a phase transition could be responsible for the band deformation induced by the decreased electron density of the quantum electron liquid. The gradual increase in $m^*$ up to ~ 3.9 $m_e$ indicates that the electron correlation becomes stronger upon K deposition (Fig. 4a), which is expected to be accomplished by lowering the density and can trigger the phase transition[15,16,28]. A series of scattering rates in the sequence of K deposition shows the drastic change of energy dependence from quadratic (#0 and #1) to linear (#2, #3, and #4) dependence (Fig. 4d,e) together with the reduction of cut-off energy, which consistently reflects the enhancement of electron correlation strength[29,30]. Linear energy dependence is a well-known behaviour of non-Fermi liquids[23,24,31], strongly suggesting that the quantum electrons transit into a distinct phase, departing from the Fermi liquid phase. Further, the deformation is temperature dependent, indicating that the observed band deformation is due to the phase transition (Extended Data Fig. 9).

According to the phase diagram of pure 2D electrons (Fig. 1a), a possible phase, accessed by reducing the density along with the strengthened electron-electron interaction, is a hexatic phase or Wigner crystal. The remaining metallic band crossing $E_F$ and the lack of a signature of translational symmetry change in the W-shaped band dispersion preclude the formation of the Wigner crystal. It is reasonable to speculate that the W-shaped band dispersion is derived from a hexatic phase as the aforementioned symmetry characteristics of the W-shaped band are coincident with those of the hexatic phase, which is a liquid crystal phase with marginally broken rotational symmetry and preserved continuous translational symmetry[28,32]. Our experimental observations thus suggest the emergence of the hexatic phase in the quantum regime and extend the phase diagram of pure 2D electrons, which can provide a further understanding of the transition process in the quantum regime.

In summary, we discovered an unprecedented 2D quantum electron liquid on the surface of a 2D electride crystal. The clear identification of the quantum electron liquid and its spin-polarized nature provide a step towards experimental accessibility of correlated electronic phases in the quantum regime. Indeed, the possible phase transition from the initial liquid phase to the phase exhibiting non-Fermi liquid behaviour, which mimics the symmetry characteristics of hexatic phase of liquid crystalline phase, was demonstrated upon the enhancement of electron correlation. We believe our results will stimulate the exploratory study of exotic phases in the quantum regime, such as the long-lasting solid phase of pure electrons that may be realized by cooling the present system and reducing the density. Furthermore, the practical feasibility of this new pure electron system – electrons on a rigid crystal surface at relatively high temperature – allows various technical approaches, not only the transport measurements but also various forms of microscopy or other spectroscopy. Indeed, as a preliminary example, we succeeded in imaging pure 2D quantum electrons in real space as well as obtaining spectroscopic evidence by scanning tunneling microscopy/spectroscopy (STM/S, Extended Data Fig. 10). Our system thus calls broad-ranging further study, which will provide new insights into intriguing quantum phenomena and extend the field to future applications such as quantum computation[33] with electrons.

## Acknowledgments

This work was supported by National Research Foundation of Korea (NRF) grant funded by the Korean government (Ministry of Science and ICT) (Grant No. 2021R1A6A1A03039696, 2022M3H4A1A01010832, No.2020R1A4A2002828, 2020K1A3A7A09080366, 2021R1A2C1013119, No.2019R1A6A3A01091336), Samsung Science and Technology Foundation under Project Number SSTF-BA2101-04, and Institute for Basic Science (Grant No. IBS-R011-D1). The Advanced Light Source is supported by the Office of Basic Energy Sciences of the US DOE under contract No.DE-AC02-05CH11231. S.H. and C.K. acknowledge support from the Institute for Basic Science in Korea (Grant No.IBS-R009-G2). Part of this study has been performed by using facilities at IBS Center for Correlated Electron Systems, Seoul National University, Korea.

## Author Contributions

S.W.K. and Y.K. conceived the project. S.Y.L, J.B. and S.W.K. synthesized and characterized single crystals. S.K., C.-Y.L., J.H., G.L., Y.L., J.D.D. and Y.K. performed ARPES measurements. S.K., J.B., S.H., C.K., and Y.K. carried out spin-resolved ARPES experiments.

S.Y.S. and J.S. conducted STM/S measurements. S.K., J.B., C.-Y.L, S.-G.K., Y.K., and S.W.K. analysed ARPES data. J.B., D.T., S.-G.K., Y.H.L., Y.K., and S.W.K. delivered DFT calculations. All authors discussed the results. S.K., J.B., Y.K. and S.W.K. prepared a manuscript with contributions from all authors.

## Competing interests

The authors declare no competing interests.

## Methods

**Single crystal growth.** Polycrystalline $[Gd_2C]^{2+} \cdot 2e^-$ rods were synthesized by the arc melting method to prepare the feed and seed materials for single crystal growth. The Gd metal and graphite pieces are mixed with a molar ratio of Gd:C = 2:1 in a glove box ($H_2O$ < 1 ppm, $O_2$ < 1 ppm) filled with purified Ar (99.999%) gas. The mixture was melted in an arc furnace under a high-purity Ar atmosphere. To obtain a single homogeneous phase, the melting process was repeated at least three times. After cooling, the polycrystalline $[Gd_2C]^{2+} \cdot 2e^-$ was shaped into the rods. For single crystal growth, the floating zone melting method was performed under a high-purity Ar atmosphere to prevent oxidation. The feed and seed rods were rotated in opposite directions at the same speed of 6 rpm, with the low melt viscosity of the $[Gd_2C]^{2+} \cdot 2e^-$ electride causing a growth speed slower than 2 mm per hour. The quality of the obtained single crystal samples was checked by a combination of X-ray diffraction (XRD) and inductively coupled plasma (ICP) spectroscopy. The clear observation of the three-fold symmetry as a rhombohedral structure in the XRD patterns of the $\phi$-scan and the negligible impurity level (all detectable impurities < 1 ppm) in the ICP results guarantee the high quality of the single crystal[19].

**ARPES measurements.** ARPES measurements were performed at beamline 4.0.3 (Merlin) of the Advanced Light Source, Lawrence Berkeley National Laboratory. Sample preparations for ARPES measurements were carried out in an Ar-filled glove box to avoid contamination from water and oxygen. Single crystals of $[Gd_2C]^{2+} \cdot 2e^-$ were attached to the sample holder for ARPES and covered by a ceramic top post to prepare for cleaving the sample in situ in the ARPES chamber. Both samples and top posts were fixed by silver epoxy. After curing the silver epoxy, the samples were immediately transferred to the UHV chamber for ARPES measurement. To obtain a clean surface, samples were cleaved in situ by UHV with a pressure better than $4 \times 10^{-11}$ Torr after cooling to 10 K. Measurements were performed at a temperature of 10 K, except temperature-dependent measurements, in which case the temperature is varied up to 40 K. Spectra were acquired with a Scienta R8000 analyser. To obtain $k_z$-dependent band dispersion, photon energies were set to the range between 80 eV and 110 eV with a 2 eV step. Resonant ARPES measurements were performed with photon energies between 143 and 149 eV with every 1 eV step around the resonant condition of the Gd $4d$ core level (148 eV). Other data, including the in-plane ($k_x$-$k_y$) Fermi surface map and band dispersions, were acquired with 90 eV of photon energy. K deposition was carried out by in situ evaporation using commercial SAES alkali metal dispensers mostly at 10 K after cleaving the electride crystal in a UHV chamber. For the temperature-dependent measurement, K deposition occurred at 40 K. The coverage of K was estimated from core-level spectra of K $3p$ and the appearance of a valence band induced by the K overlayer (Extended Data Fig. 4). The total energy resolution was 25 meV or better.

**Spin-resolved ARPES measurements.** Spin-resolved ARPES measurements were performed at the Center for Correlated Electron Systems, Institute for Basic Science. Similar to ARPES measurements, sample preparations for spin-resolved ARPES measurements were carried out in an Ar-filled glove box to avoid contamination from water and oxygen and immediately transferred to the UHV chamber for spin-resolved ARPES measurement. The clean surface was obtained by cleaving the sample in situ in the UHV with a pressure better than $5.0 \times 10^{-11}$ Torr after cooling to 10 K. Measurements were performed at the temperature of 10 K. Spectra were acquired with a Phoibos 225 analyser with VLEED-type spin detectors. The photon energy was 21.2 eV (He I). The total energy resolution was approximately 50 meV. The spin-resolved EDCs ($I_{up}$ and $I_{down}$) in Fig. 3f–i were obtained from the observed spectra for different target magnetization ($I_{+M}$ and $I_{-M}$) by the following relation,

$$I_{up} = (1 + P) \ (I_{+M} + I_{-M})/2, \ I_{down} = (1 - P) \ (I_{+M} + I_{-M})/2$$

where $P$ is spin polarisation which can be estimated by $P = (I_{+M} - I_{-M})/\{(I_{+M} + I_{-M}) \ S_{eff}\}$. $S_{eff}$ is an effective Sherman function with a value of ~ 0.3 in the present work.

**STM/S measurements.** STM experiments were performed using a home-built low-temperature STM at Department of Emerging Materials Science, DGIST. The single crystal of $[Gd_2C]^{2+}\cdot2e^-$ was cleaved at 20 K in a UHV with a pressure better than $4 \times 10^{-11}$ Torr and immediately plugged into the STM head. A mechanically sharpened PtIr wire was used for an STM tip. To acquire the differential conductance ($dI/dV$) spectra and $dI/dV$ maps, we used a standard lock-in technique with a modulation frequency $f = 718$ Hz and root-mean-square amplitude $V_{mod} = 5$ mV. Measurements were performed at a temperature of 4.3 K.

**Self-energy analysis.** To obtain the information of the imaginary part of the self-energy, EDCs were fitted as the ARPES intensity, which is nothing but the spectral function with the following form:

$$A(\mathbf{k}, \omega) = -\frac{1}{\pi}\frac{Im\Sigma}{(\omega - \varepsilon_{\mathbf{k}} - Re\Sigma)^2 + Im\Sigma^2}$$

where $Re\Sigma$ is the real part of the self-energy and $\varepsilon_{\mathbf{k}}$ is the bare band dispersion. The obtained half-width at half-maximum of the Lorentzian curve is the imaginary part of the self-energy ($Im\Sigma$). Prior to fitting, EDCs were divided by Fermi-Dirac distribution to avoid diminishing intensity near the Fermi level. The fitting curve involves two Lorentzian curves: lower and higher binding energies for surface electron and IAE bands, respectively. The fitted curve was convoluted with a Gaussian to encounter the experimental resolution. The resulting fit curves, multiplied by the Fermi-Dirac distribution, closely reproduce the original spectra (Supplementary Figure 3), reflected in small fitting error of $Im\Sigma$ in Figs. 3c and 4d.

**Effective mass estimation**. The values of effective mass $m^*$ in Fig. 4a were estimated by Fermi velocity ($v_F$) obtained at each K deposition step based on the following relation:

$$m^* = \frac{\hbar k_F}{v_F} = \hbar^2 k_F / \frac{dE(\mathbf{k})}{d\mathbf{k}}$$

where $\hbar$ is the Planck constant and $k_F$ is the Fermi momentum. Supplementary Figure 4 shows the peak positions of the band dispersion near $E_F$ at each deposition step and the fitted curve of the linear function. $v_F$ was obtained from the slope of the linear fit curve, $\frac{dE(\mathbf{k})}{d\mathbf{k}}$, and $m^*$ of each step was estimated from the above equation using the obtained $v_F$. In particular, for the pristine sample (deposition step #0 of Fig. 4a), $m^*$ was double-checked using parabolic band fitting (Fig. 3b), which matches well with the estimated $m^*$ value from $v_F$.

**Autocorrelation analysis.** Autocorrelation of ARPES intensity $I_{AC}(\mathbf{q}, \omega)$ (Extended Data Fig. 5e) was calculated by substituting the symmetrized and filtered CECs into the following formula[34,35]:

$$I_{AC}(\mathbf{q}, \omega) = \int I(\mathbf{k}, \omega)I(\mathbf{k} + \mathbf{q}, \omega)d\mathbf{k}$$

where $I(\mathbf{k}, \omega)$ is the ARPES intensity at momentum $\mathbf{k}$ and binding energy $\omega$ extracted by a given CEC, and $\mathbf{q}$ is the momentum transfer. The three-fold symmetrisation was applied to avoid the momentum-dependent modulation of intensity due to the matrix element effect. It was ensured that the symmetrisation does not deform the original shape of CECs from the surface band (Extended Data Fig. 5c,d). The signal from the outer bands was filtered out to include only the signal from the surface electron band in the analysis (Extended Data Fig. 5d).

**DFT calculations.** DFT calculations were performed using the generalized gradient approximation (GGA) with the Perdew-Burke-Ernzerhof (PBE) functional and the projector augmented plane-wave method implemented in the Vienna *ab initio* simulation program (VASP) code[36–38]. The 4*f*, 5*s*, 5*p*, 5*d*, and 6*s* electrons of Gd and the 2*s* and 2*p* electrons of C were used as valence electrons. The plane-wave-basis cut-off energy was set to 600 eV. The slab supercell of $a \times b \times 3c$ (27 atoms) with a vacuum layer of 20 Å along the *c*-axis was used for the surface calculation. Structural relaxations of the bulk and slab structures were performed using $8 \times 8 \times 2$ and $8 \times 8 \times 1$ *k*-point meshes until the Hellmann-Feynman forces were less than $10^{-5}$ and $10^{-3}$ eV · Å$^{-1}$, respectively. The lattice constants of the relaxed bulk structure are well matched with those obtained by the XRD measurement (mismatch of the lattice parameters is less than 1%, Extended Data Fig. 1d). For the relaxation of the slab supercell, the three layers at both ends of the slab were relaxed while keeping the central three layers frozen (Extended Data Fig. 2a). The empty spheres with Wigner–Seitz radius of 1.25 Å were placed both on the surface and at the interlayer space to obtain the projected density of states of the surface and interlayer positions, respectively. We performed the ELF analysis formulated by Becke and Edgecombe[39], which is appropriate for investigating the spatial distribution of electrons devoid of atomic orbitals in the electride system[40]. The crystal structures and ELF were visualized with the VESTA code[41]. The Fermi surface was visualized with the XCrySDen code[42].

# References

1. Phillips, P., Wan, Y., Martin, I., Knysh, S. & Dalidovich, D. Superconductivity in a two-dimensional electron gas. *Nature* **395**, 253–257 (1998).

2. Mannhart, J. & Schlom, D. G. Oxide interfaces—an opportunity for electronics. *Science* **327**, 1607–1611 (2010).

3. Hwang, H. Y. *et al.* Emergent phenomena at oxide interfaces. *Nat. Mater.* **11**, 103–113 (2012).

4. Wigner, E. On the interaction of electrons in metals. *Phys. Rev.* **46**, 1002–1011 (1934).

5. Sommer, W. T. & Tanner, D. J. Mobility of electrons on the surface of liquid $^4$He. *Phys. Rev. Lett.* **27**, 1345–1349 (1971).

6. Grimes, C. C., Brown, T. R., Burns, M. L. & Zipfel, C. L. Spectroscopy of electrons in image-potential-induced surface states outside liquid helium. *Phys. Rev. B* **13**, 140–147 (1976).

7. Grimes, C. C. Electrons in surface states on liquid helium. *Surf. Sci.* **73**, 379–395 (1978).

8. Grimes, C. C. & Adams, G. Evidence for a liquid-to-crystal phase transition in a classical two-dimensional sheet of electrons. *Phys. Rev. Lett.* **42**, 795–798 (1979).

9. Grimes, C. C. & Adams, G. Crystallization of electrons on the surface of liquid helium. *Surf. Sci.* **98**, 1–7 (1980).

10. Andrei, E. Y. *Two-dimensional electron systems on helium and other cryogenic substrates*. (Kluwer Academic Publishers, 1997).

11. Monarkha, Y. & Kono, K. *Two-dimensional coulomb liquids and solids*. (Springer, 2004).

12. Gallet, F., Deville, G., Valdes, A. & Williams, F. I. B. Fluctuations and shear modulus of a classical two-dimensional electron solid: Experiment. *Phys. Rev. Lett.* **49**, 212–215 (1982).

13. Deville, G., Valdes, A., Andrei, E. Y. & Williams, F. I. B. Propagation of shear in a two-dimensional electron solid. *Phys. Rev. Lett.* **53**, 588–591 (1984).

14. Marty, D. Stability of two-dimensional electrons on a fractionated helium surface. *J. Phys. C: Solid State. Phys.* **19**, 6097–6104 (1986).

15. Spivak, B. & Kivelson, S. A. Phases intermediate between a two-dimensional electron liquid and Wigner crystal. *Phys. Rev. B* **70**, 155114 (2004).

16. Waintal, X. On the quantum melting of the two-dimensional Wigner crystal. *Phys. Rev. B* **73**, 075417 (2006).

17. Lee, K., Kim, S. W., Toda, Y., Matsuishi, S. & Hosono, H. Dicalcium nitride as a two-dimensional electride with an anionic electron layer. *Nature* **494**, 336–340 (2013).

18. Park, J. *et al.* Strong localization of anionic electrons at interlayer for electrical and magnetic anisotropy in two-dimensional $Y_2C$ electride. *J. Am. Chem. Soc.* **139**, 615–618 (2017).

19. Lee, S. Y. *et al*. Ferromagnetic quasi-atomic electrons in two-dimensional electride. *Nat. Commun.* **11**, 1526 (2020).

20. Santander-Syro, A. F. *et al.* Two-dimensional electron gas with universal subbands at the surface of $SrTiO_3$. *Nature* **469**, 189–193 (2011).

21. Meevasana, W. *et al.* Creation and control of a two-dimensional electron liquid at the bare $SrTiO_3$ surface. *Nat. Mater.* **10**, 114–118 (2011).

22. Wang, Z. *et al.* Tailoring the nature and strength of electron–phonon interactions in the $SrTiO_3$(001) 2D electron liquid. *Nat. Mater.* **15**, 835–839 (2016).

23. Damascelli, A., Hussain, Z. & Shen, Z. -X. Angle-resolved photoemission studies of the cuprate superconductors. *Rev. Mod. Phys.* **75**, 473–541 (2003).

24. Kaminski, A. & Fretwell, H. M. On the extraction of the self-energy from angle-resolved photoemission spectroscopy. *New. J. Phys.* **7**, 98 (2005).

25. Ceperley, D. Ground state of the fermion one-component plasma: A Monte Carlo study in two and three dimensions. *Phys. Rev. B* **18**, 3126–3138 (1978).

26. Varsano, D., Moroni, S., & Senatore, G. Spin-polarization transition in the two-dimensional electron gas. *Europhys. Lett.* **53**, 348–353 (2001).

27. Attaccalite, C., Moroni, S., Gori-Giorgi, P. & Bachelet, G. B. Correlation energy and spin polarization in the 2D electron gas. *Phys. Rev. Lett.* **88**, 256601 (2002).

28. Clark, B. K., Casula, M. & Ceperley, D. M. Hexatic and mesoscopic phases in a 2D quantum Coulomb system. *Phys. Rev. Lett.* **103**, 055701 (2009).

29. Miyake, K., Matsuura, T. & Varma, C. M. Relation between resistivity and effective mass in heavy-fermion and A15 compounds. *Solid State Commun.* **71**, 1149–1153 (1989).

30. Jacko, A. C., Fjærestad, J. O. & Powell, B. J. A unified explanation of the Kadowaki-Woods ratio in strongly correlated metals. *Nat. Phys.* **5**, 422–425 (2009).

31. Varma, C. M., Littlewood, P. B., Schmitt-Rink, S., Abrahams, E. & Ruckenstein, A. E. Phenomenology of the normal state of Cu-O high-temperature superconductors. *Phys. Rev. Lett.* **63**, 1996–1999 (1989).

32. Halperin, B. I. & Nelson, D. R. Theory of two-dimensional melting. *Phys. Rev. Lett.* **41**, 121–124 (1978).

33. Platzman, P. M. & Dykman, M. I. Quantum computing with electrons floating on liquid helium. *Science* **284**, 1967 (1999).

34. Borisenko, S. V. *et al.* Pseudogap and charge density waves in two dimensions. *Phys. Rev. Lett.* **100**, 196402 (2008).

35. Hashimoto, M. *et al.* Reaffirming the $d_{x^2-y^2}$ superconducting gap using the autocorrelation

angle-resolved photoemission spectroscopy of $Bi_{1.5}Pb_{0.55}Sr_{1.6}La_{0.4}CuO_{6+\delta}$. *Phys. Rev. Lett.* **106**, 167003 (2011).

36. Kresse, G. & Furthmüller, J. Efficient iterative schemes for *ab initio* total-energy calculations using a plane-wave basis set. *Phys. Rev. B* **54**, 11169–11186 (1996).

37. Blöchl, P. E. Projector augmented-wave method. *Phys. Rev. B* **50**, 17953–17979 (1994).

38. Perdew, J. P., Burke, K. & Ernzerhof, M. Generalized gradient approximation made simple. *Phys. Rev. Lett.* **77**, 3865–3868 (1996).

39. Becke, A. D. & Edgecombe, K. E. A simple measure of electron localization in atomic and molecular systems. *J. Chem. Phys.* **92**, 5397–5403 (1990).

40. Scanlon, D. O. Leading the charge of electride discovery. *Matter* **1**, 1113–1114 (2019).

41. Momma, K. & Izumi, F. *VESTA 3* for three-dimensional visualization of crystal, volumetric and morphology data. *J. Appl. Crystallogr.* **44**, 1272–1276 (2011).

42. Kokalj, A. XCrySDen—a new program for displaying crystalline structures and electron densities. *J. Mol. Graph. Model.* **17**, 176–179 (1999).

## Data Availability

All data supporting the findings of this work are included in the main text, extended data, and supplementary information. These are available from the corresponding authors upon reasonable request.

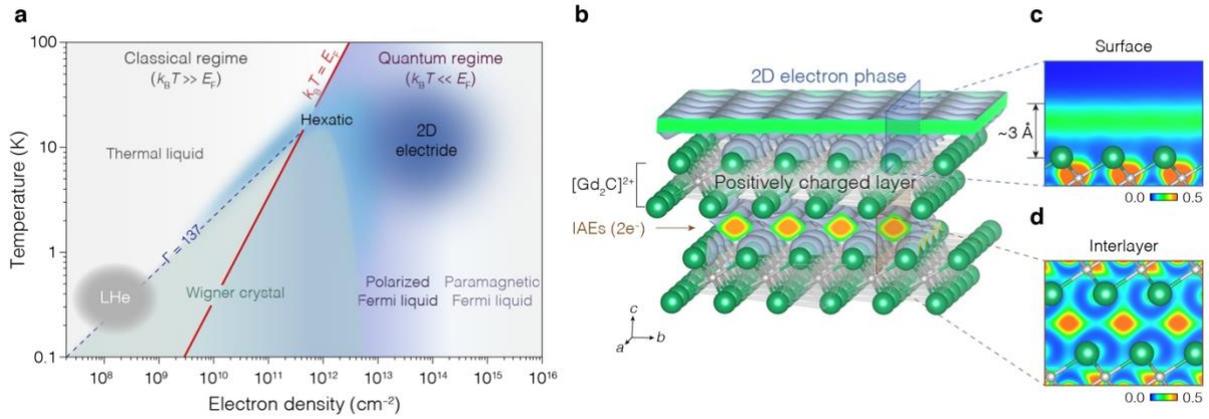

**Figure 1 | Evolution of pure 2D quantum electrons. a**, Phase diagram of the pure 2D electron system in terms of electron density and temperature. The solid red line ($k_B T = E_F$), where kB is Boltzmann constant and T is temperature, divides the quantum (right) from classical (left) regimes. Colours represent the electron phases; the thermal liquid, hexatic liquid crystal and Wigner crystal in the classical regime; and paramagnetic Fermi liquid, polarized Fermi liquid, hexatic liquid crystal and Wigner crystal in the quantum regime. The phase boundaries are estimated from the literature[8, 25–28]. The electron phases on the LHe (grey area) are taken from the literature[8]. The dashed blue line is experimental estimation of Coulomb coupling parameter, $\Gamma$ = 137 for the Wigner crystallization in the classical regime[8]. The present pure 2D quantum electrons on the 2D [Gd₂C]²⁺·2e⁻ electride reside in the area of Fermi liquid (dark blue area). Detailed physical quantities of 2D electrons on the surface of LHe and 2D [Gd₂C]²⁺·2e⁻ electride are compared in Supplementary Table 1. **b–d,** Visualization of ELFs for the electrons on the surface and at the intralayer of 2D [Gd₂C]²⁺·2e⁻ electride. Cross-sectional plots of the ELFs for the electrons on the surface (**c**) and IAEs at the interlayer (**d**) taken from the blue and brown windows marked in **b**, respectively. The colour bars in **c** and **d** represent ELF.

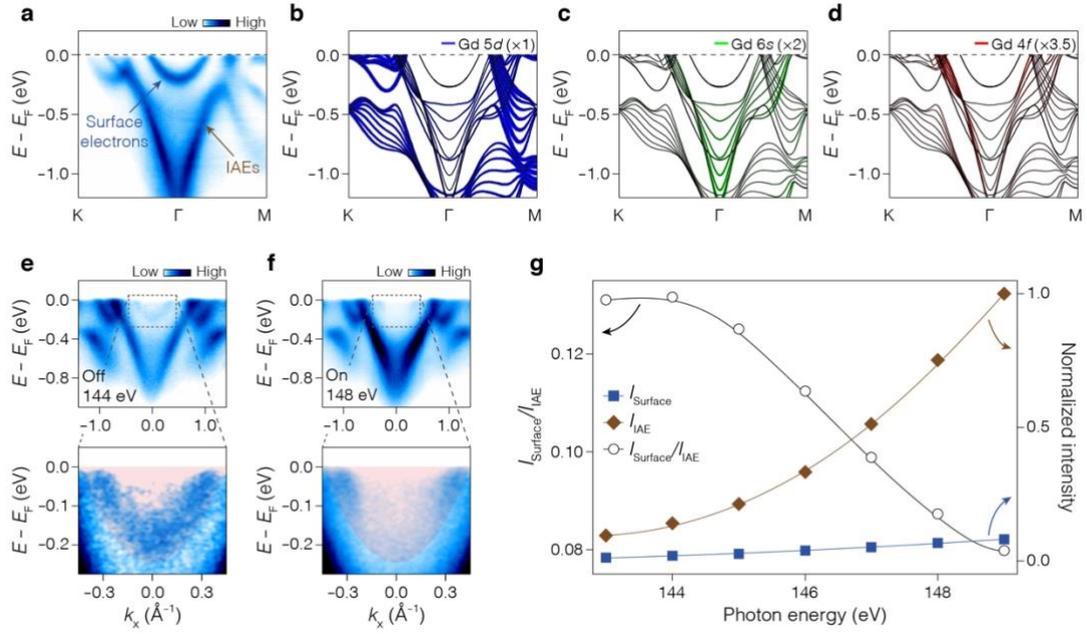

**Figure 2 | Floating quantum electrons on the 2D electride. a**, Band dispersion of the 2D [Gd$_2$C]$^{2+}$·2e$^-$ observed by ARPES. The blue and brown arrows indicate the parabolic and V-shaped band dispersion, respectively, from the surface electrons on the terminated [Gd$_2$C]$^{2+}$ layer and IAEs trapped between [Gd$_2$C]$^{2+}$ layers in the bulk, respectively (Supplementary Figure 1). **b–d**, Calculated band dispersions with fat band analysis using the nine-slab model (Extended Data Fig. 2a), exhibiting the contribution of Gd 5$d$ (**b**), 6$s$ (**c**), and 4$f$ (**d**) orbitals. Band thickness with different colours represents the contribution of each orbital; the thicker line reflects a greater contribution from the corresponding orbitals. The fat band width in **c** and **d** has been magnified by 2 and 3.5 times, respectively, compared with that in **b**. Almost no colour from each orbital is represented in the parabolic band. **e,f**, Band dispersions taken at off-resonant (144 eV, **e**) and on-resonant (148 eV, **f**) conditions of the Gd 4$d$ core level. Lower panels are magnified images of the boxed area in the upper panels. **g**, Normalized intensity (right) of the surface 2D electrons ($I_{Surface}$, blue) and IAEs ($I_{IAE}$, brown), and their ratio (left, black) as a function of photon energy (Extended Data Fig. 3d,e). $I_{Surface}$ at each photon energy is obtained in the upper panels of **e** and **f** by integrating the intensity of the shaded red area in the lower panels, and $I_{IAE}$ at each photon energy is obtained in the upper panels of **e** and **f** by integrating the intensity of the remaining are except the area for $I_{Surface}$. The colour bars in **a**, **e**, and **f** represent ARPES intensity in arbitrary units.

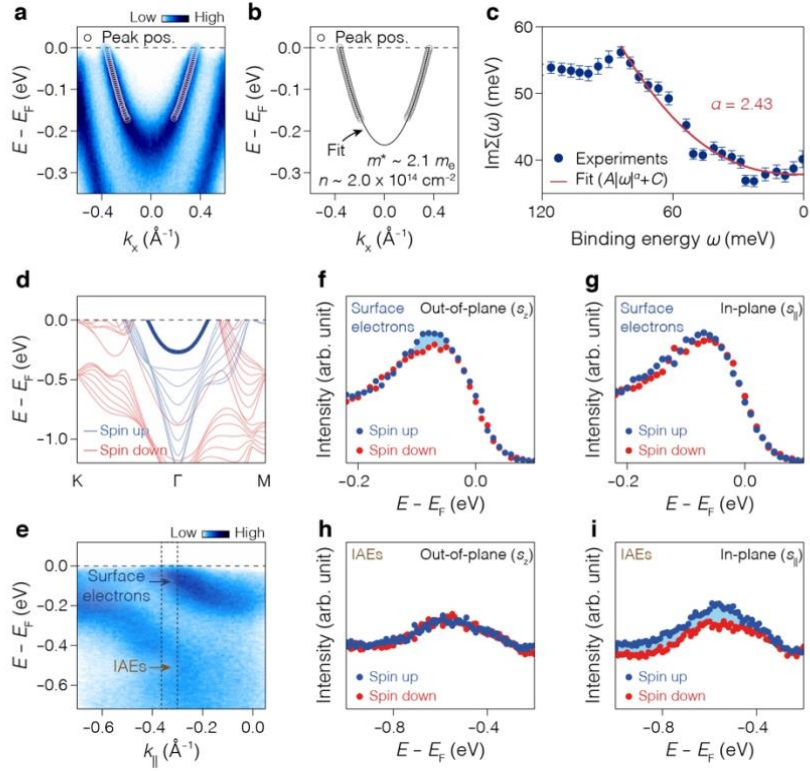

**Figure 3 | Spin-polarized nature of quantum electron liquid. a,b,** Enlarged band dispersion of surface electrons (**a**); the $m^*$ and $n$ of surface electrons are estimated to be ~ 2.1$m_e$ and ~ 2.0 $\times 10^{14}$ cm$^{-2}$, respectively, from the quadratic fitting (black curve in **b**) of peak positions (white circles) obtained from momentum distribution curves (MDCs) in **a**. **c,** Imaginary part of the self-energy, Im$\Sigma$ (filled blue circles), obtained by fitting the energy distribution curve (EDC) width (Supplementary Figure 3) as a function of binding energy ($\omega$). Error bar represents the fitting error of each point. The solid red curve represents the fit with a power law (Im$\Sigma(\omega)$ = A$|\omega|^\alpha$ + C). A, C, and $\alpha$ are fitting parameters. **d,** Calculated spin-polarized band structure with a fat band for the surface electrons. Blue and red colours reflect the spin up and spin down components, respectively. A thick blue band for the surface electrons clearly indicates that the quantum electrons are spin-polarized. For the V-shaped bands of the IAEs, spin-polarized state is also shown by the blue colour. **e,** Measured band dispersion for the spin-resolved ARPES. **f–i,** Spin-resolved EDCs of surface electrons (**f** and **g**) and IAEs (**h** and **i**); $s_\perp$ and $s_\parallel$ represent spins with two different directions perpendicular and parallel to the plane, respectively; a.u., arbitrary units. Spin-up and spin-down components are represented by blue and red dots, respectively. EDCs are obtained by integrating the intensity within the dashed vertical lines in **e**. The area shaded by light blue in **f** and **i** represent the difference between spin-up and spin-down. The colour bars in a and e represent ARPES intensity in arbitrary units.

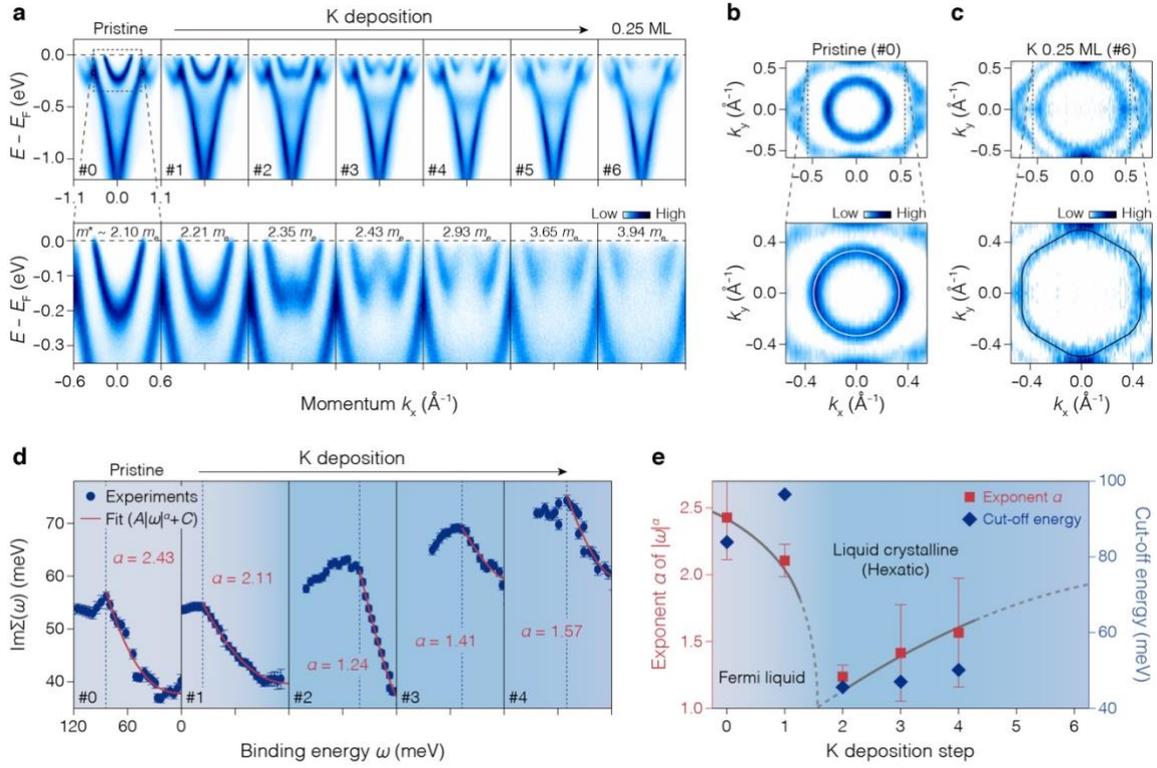

**Figure 4 | Phase transition of quantum electrons. a**, Stepwise evolution of the electronic structure upon K deposition, from 0 ML (pristine, leftmost; ML, monolayer) up to 0.25 ML coverage (rightmost). Details of K coverage estimation are described in Extended Data Fig. 4. Lower panels show magnified images (dashed box in the leftmost upper panel) for the band of quantum electrons in each step. The values of effective mass were obtained from the band slope near $E_F$ (see Methods and Supplementary Figure 4). **b,c**, Fermi surface in the $k_x$–$k_y$ plane of the as-cleaved (#0, **b**) and 0.25 ML K-deposited (#6, **c**) samples. Lower panels exhibit magnified images of dashed boxes in upper panels. Overlaid solid lines on magnified images represent guides for the Fermi surface topology. The colour bars in **a–c** represent ARPES intensity in arbitrary units. **d**, The Im$\Sigma$ for the band of quantum electrons upon the K deposition step, extracted by the line shape analysis of EDCs (Supplementary Figure 3). Solid red curves in each panel are fitted curves of Im$\Sigma(\omega)$ with a power law. Dashed blue vertical lines indicate the cut-off energy of Im$\Sigma$. Error bars represent the fitting error of each point. **e**, Extracted quantities of exponent $\alpha$ in Im$\Sigma(\omega) = A|\omega|^\alpha + C$ and the cut-off energy as a function of the K deposition. Error bars of exponent $\alpha$ represent the fitting error of each point. Both the exponent $\alpha$ (red squares) and the cut-off energy (blue diamonds) show a similar trend, having minima at a specific deposition step (#2) and then increasing towards the initial values upon K coverage. The grey curve serves as a guide for the eye.

**Extended data figures and tables**

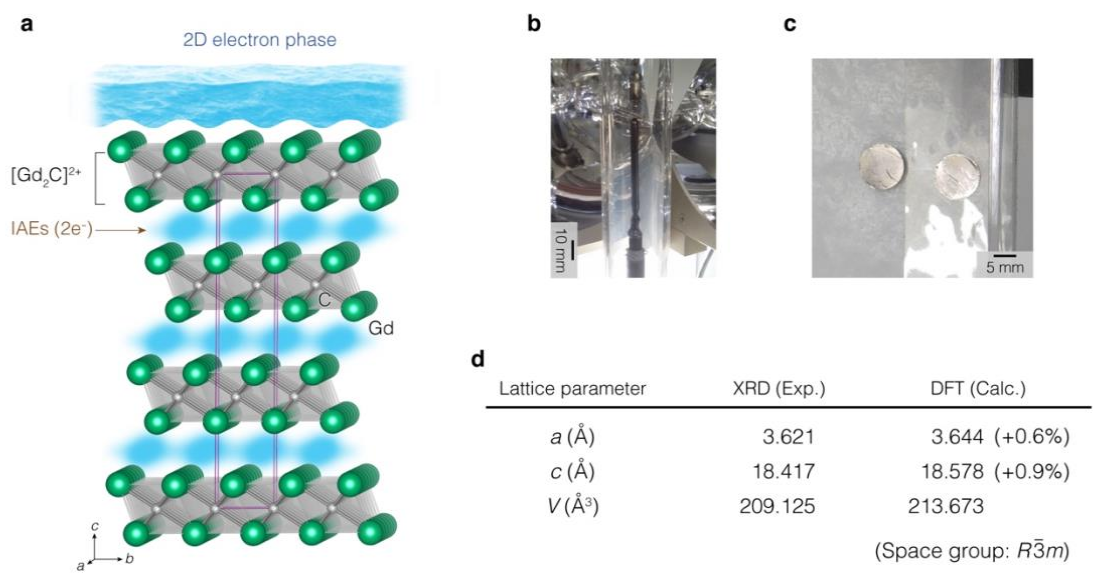

| d Lattice parameter | XRD (Exp.) | DFT (Calc.) |
|---|---|---|
| $a$ (Å) | 3.621 | 3.644 (+0.6%) |
| $c$ (Å) | 18.417 | 18.578 (+0.9%) |
| $V$ (Å³) | 209.125 | 213.673 |

(Space group: $R\bar{3}m$)

**Extended Data Figure 1 | Crystal structure of the 2D [Gd₂C]²⁺·2e⁻ electride. a,** Schematic illustration of the 2D quantum electron liquid on the surface of [Gd₂C]²⁺·2e⁻ electride. It has anti-CdCl₂-type layered structure with a space group of $R\bar{3}m$. The cationic slab [Gd₂C]²⁺ is composed of an edge-sharing octahedra structure and is separated by the interlayer space. IAEs (brown arrow) are confined between the positively charged [Gd₂C]²⁺ layers. The 2D electrons distinct from IAEs are floated on the topmost [Gd₂C]²⁺ layer. The purple lines indicate the unit cell. **b,c,** Photographs of single crystal [Gd₂C]²⁺·2e⁻ rod with 50 mm length grown by floating zone melting method (**b**) and as-cleaved crystal by 3M Scotch tape (**c**). **d,** Lattice parameters obtained by the XRD measurements[19] and the DFT calculations.

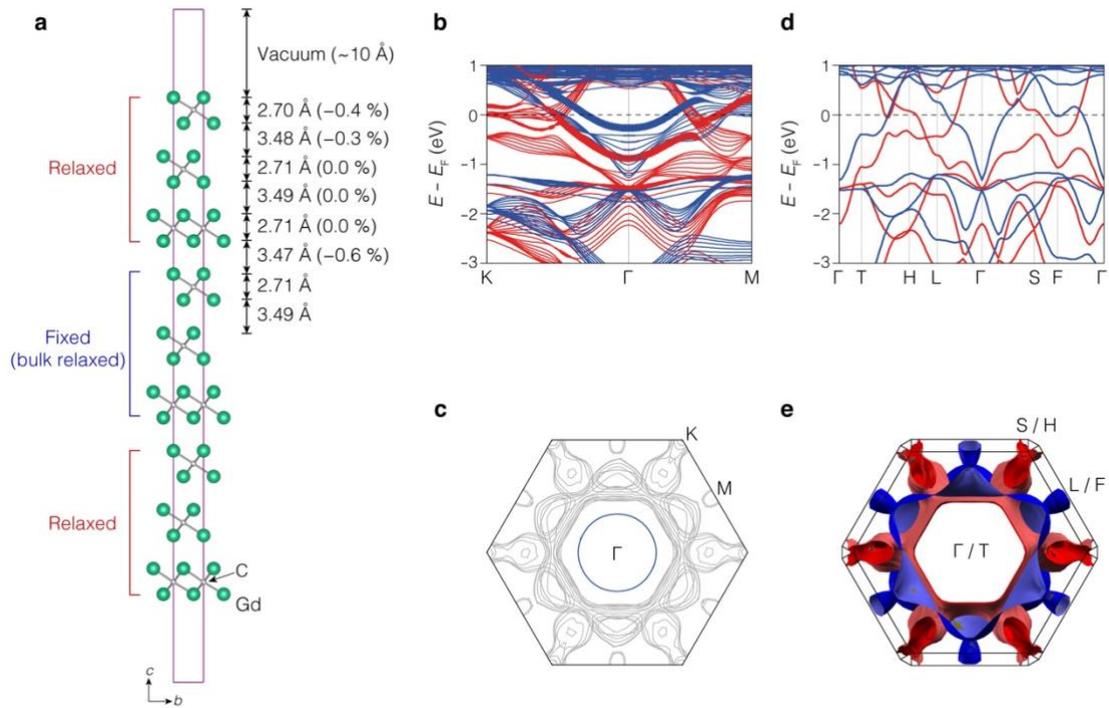

**Extended Data Figure 2 | The detailed electronic structure of [Gd$_2$C]$^{2+}$·2e$^-$ obtained by DFT calculations. a**, Relaxed crystal structure of the nine-slab model for the slab calculations. $a \times b \times 3c$ supercell with a vacuum of 20 Å along the $c$-axis was used. For the relaxation of the nine-slab model, the optimised bulk structure was used for the central three-layer, while the three layers at both ends of the slab were relaxed. **b,c**, Calculated band structure and the Fermi surface of [Gd$_2$C]$^{2+}$·2e$^-$ using the nine-slab model, respectively. Blue and red colours indicate spin up and spin down components, respectively. Band thickness represents the contribution of the surface electrons. The thicker line reflects a contribution mainly from the surface character. **d,e**, Band structure and the Fermi surface (top view along $k_z$ direction) of [Gd$_2$C]$^{2+}$·2e$^-$ obtained by bulk calculation, respectively. In contrast to the result from the nine-slab model, the bulk calculation does not reproduce the parabolic band dispersion with cylindrical 2D Fermi surface observed by ARPES (Fig. 2a and Extended Data Fig. 3a–c).

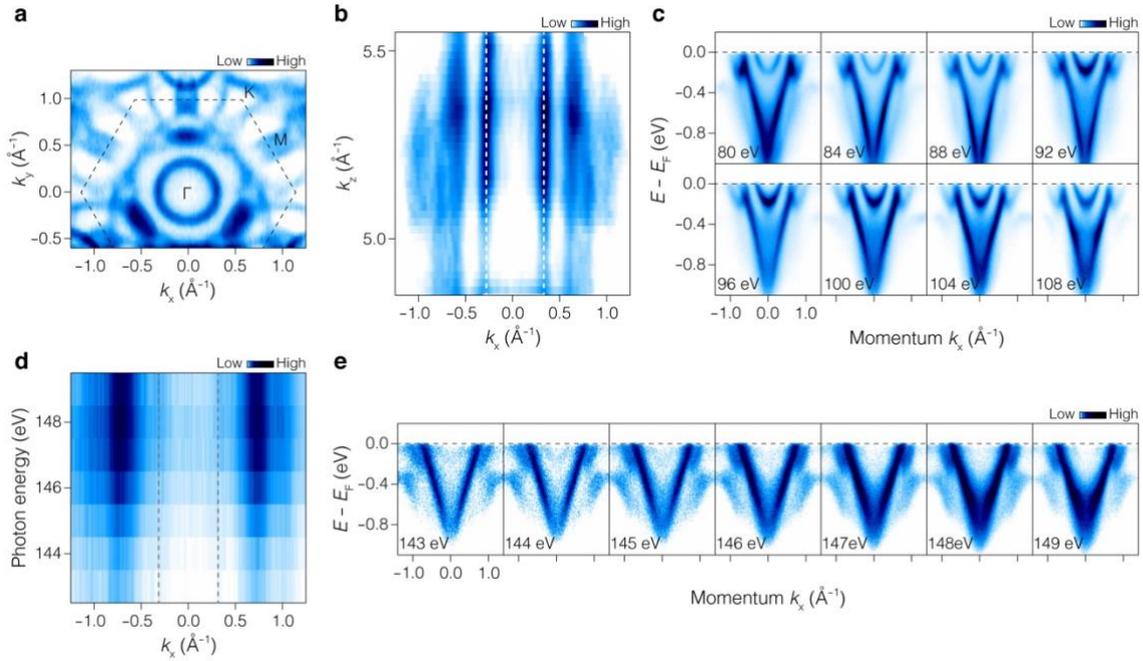

**Extended Data Figure 3 | Fermi surface of surface electrons on [Gd₂C]²⁺·2e⁻ and photon energy dependence. a**, In-plane ($k_x$–$k_y$) Fermi surface of [Gd₂C]²⁺·2e⁻. The black dashed line indicates the BZ. Circular Fermi surface near BZ centre corresponds to the parabolic band dispersion exhibited in Fig. 2a. **b**, Fermi surface in the $k_x$–$k_z$ plane. The inner straight surface, overlaid by dashed white lines, is from the parabolic band shown in Fig. 2a and from the circular Fermi surface in **a**. **c**, Band dispersion observed at various photon energies from 80 to 108 eV with every 4 eV step to obtain $k_x$–$k_z$ plane Fermi surface in **b**. **d**, Photon energy-dependent ARPES intensity map extracted from the Fermi level taken with photon energies from 143 to 149 eV, which includes both on- and off-resonant conditions of Gd 4$d$ core level. Dashed grey lines are overlaid on the surface 2D electron band as a guide for the eye. The intensity of IAE band at 148 eV greatly increased, a clear resonance behaviour while changing in the intensity of the surface 2D electron band is far weak through the whole photon energies. **e**, The detailed band dispersions at each photon energy in **d** are exhibited.

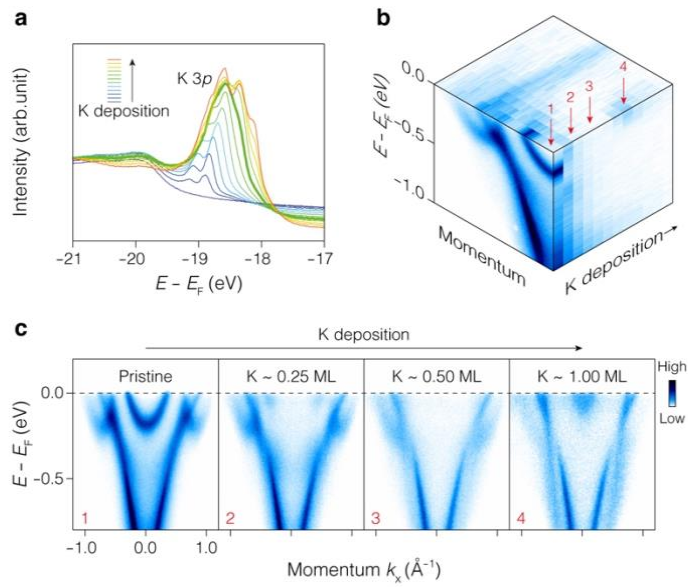

**Extended Data Figure 4 | Determination of K coverage. a**, Core-level spectra of K $3p$ with different K coverage on the cleaved $[Gd_2C]^{2+} \cdot 2e^-$ surface. The K $3p$ core-level peak starts to grow with K deposition near the binding energy of 19 eV and eventually saturates, where we estimate the coverage as 1 ML (thick green curve). Above 1 ML, chemically shifted additional peaks emerge at lower binding energy close to 18 eV. **b**, A 3D representation of the band evolution with increasing K coverage. **c**, Extracted band dispersion at several different K coverages indicated by red arrows in **b**. Surface 2D electron band (1) evolves via K deposition (2) and disappears (3). The complete disappearance of the electron band implies that the surface electron density is actually reduced by K deposition. In the rightmost panel (4), corresponding to K 1 ML, which we estimate with core-level spectra in **a**, K band appears near Fermi level, which well agrees with the core level estimation.

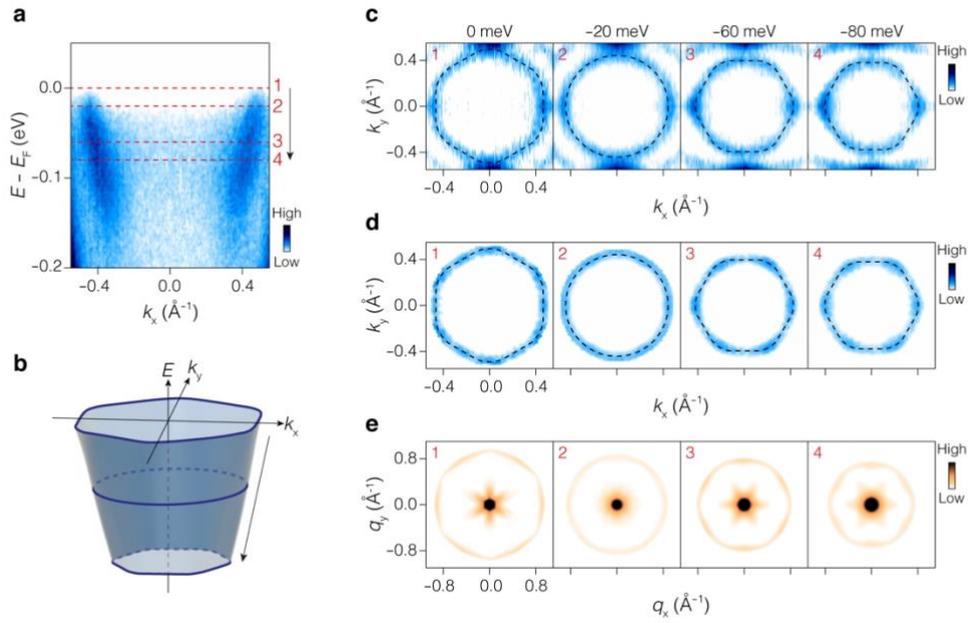

**Extended Data Figure 5 | Analysis for the hexagonal topology. a**, Zoomed-in band dispersion of 0.25 ML K coverage, corresponding to deposition step #6 in Fig. 4a. **b**, Schematic band structure of the ARPES result in **a**. Blue solid lines schematically show constant energy contours (CECs) at different energies. **c**, CECs obtained at binding energies of 0, 20, 60, and 80 meV, indicated with dashed red lines (1–4) in **a**. **d**, Symmetrised CECs with 3-fold rotation after removing the intensity from the outer band in **c**, which reveals the shape of energy contours from the quantum electrons more clearly. Overlaid dashed black lines in **c** and **d** are guides for the topology of each CEC. The guidelines evidence that the original shape of contours is not deformed by 3-fold symmetrisation. **e**, Autocorrelation analyses performed with CECs of corresponding energies in **d**, which clearly visualise the angular dependence of the band along with the in-plane momentum.

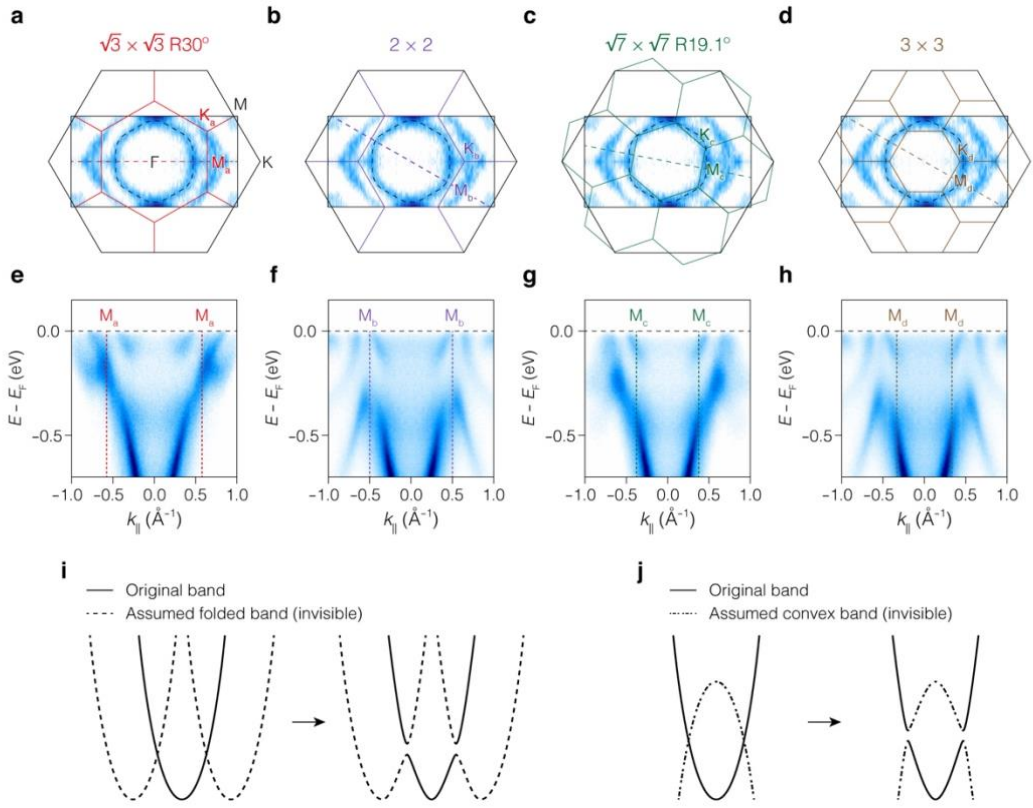

**Extended Data Figure 6 | Scenarios of W-shape band deformation. a–d**, Fermi surface with 0.25 ML K coverage. BZs of several possible surface superstructures with sub-ML coverage of K, $\sqrt{3} \times \sqrt{3}\ R30°$ (**a**), $2 \times 2$ (**b**), $\sqrt{7} \times \sqrt{7}\ R19.1°$ (**c**), and $3 \times 3$ (**d**), respectively, are overlaid with a different colour. The outermost hexagon with solid black line represents the BZ of $1 \times 1$ unit cell of $[Gd_2C]^{2+}\cdot 2e^-$. **e–h**, Observed band dispersion along the $\Gamma$–$M_{a,b,c,d}$ direction for new BZs of assumed superstructures at 0.25 ML K coverage. The new zone boundaries by assumed K orderings in **a–d** are overlaid (dashed lines with corresponding colours). Absence of W-shape band at higher momentum above the assumed new zone boundaries, which can be induced by band folding, evidences that W-shaped band deformation is not due to the ordering of deposited K. **i,j**, Schematic drawings of conceivable scenarios for band deformation (hybridisation) by K deposition with assumed zone folded band (**i**) and upward convex band (**j**), respectively. Solid and dashed bands show the original and assumed bands, respectively. Both scenarios may exhibit band deformation; however, they should form the fragment of the original electron band at higher binding energy near BZ centre, which is absent in ARPES results (Fig. 4a).

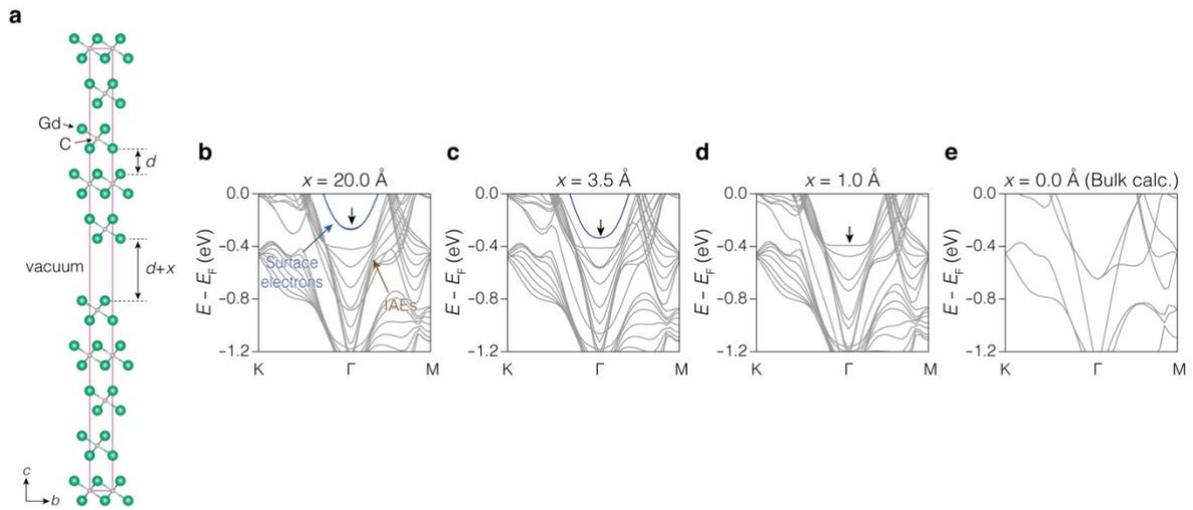

**Extended Data Figure 7 | Band evolution induced by underlying lattice potential. a**, Crystal structure of the nine-slab model for the slab calculations. The vacuum layer with the thickness $x$ along the $c$-axis is introduced in addition to the original interlayer space ($d$). **b−e**, Calculated band structures depending on the vacuum thickness $x$ of 20.0, 3.5, 1.0 and 0.0 Å, where the result of 0.0 Å is obtained by bulk calculation. Black arrows indicate the gradual shift of the surface electron band (blue) downward to the higher binding energy with a decrease in the vacuum thickness $x$. This exhibits that the surface electron state follows V-shape IAE band when it is strongly affected by underlying lattice.

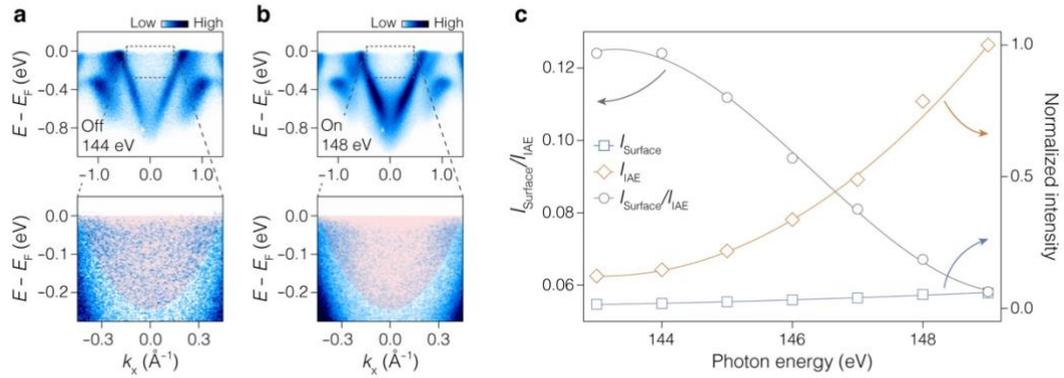

**Extended Data Figure 8 | Resonant ARPES measurement after K deposition. a**,**b**, Band dispersions taken at off- (144 eV, **a**) and on-resonant (148 eV, **b**) condition of Gd 4$d$ core level after K deposition (deposition step #6). Lower panels are magnified images of the enclosed area by black dashed box in the upper panels. **c**, Normalised intensity (right) of the surface 2D electrons ($I_{Surface}$, blue) and trapped IAEs ($I_{IAE}$, brown), and their ratio (left, gray) as a function of photon energy. $I_{Surface}$ and $I_{IAE}$ are obtained by integrating the intensity of the shaded red area in lower panels and by integrating the remaining are except for $I_{Surface}$ in upper panels of **a** and **b**, respectively.

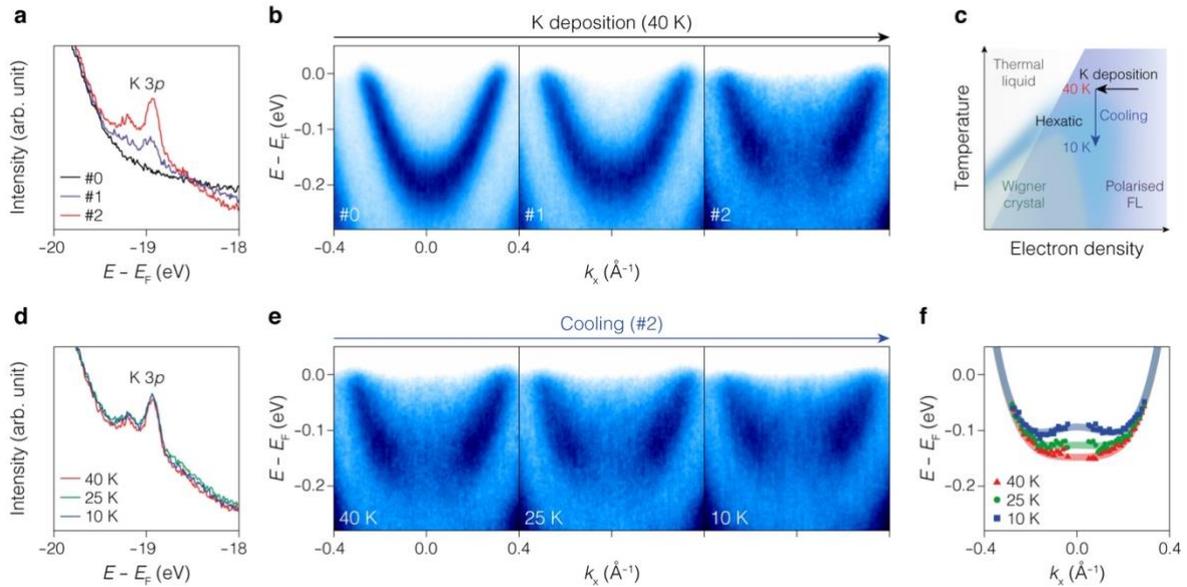

**Extended Data Figure 9 | Phase transition via decreasing temperature. a**, Core-level spectra of K 3$p$ with different K coverage at 40 K. **b**, Surface electron band dispersion with respect to K deposition. Corresponding deposition steps to the core-level spectra in **a** are indicated in the panel (#0–#2). Despite the broadening of the spectra and the slight reduction of the band minimum energy after the K deposition, the entire band dispersion still preserves nearly parabolic close to the pristine case. **c**, Enlarged phase diagram of pure 2D electron phase taken from Fig. 1a. ARPES measurements were performed by following the process marked in the phase diagram by black and blue arrows. **d**, Core-level spectra of K 3$p$ observed at different temperatures. The preserved intensity of the K 3$p$ core level indicates the absence of K desorption during the cooling process. **e**, Temperature-dependent surface electron band dispersion observed after K deposition (deposition step #2). Temperatures were set to 40, 25, and 10 K. **f**, Peak positions (markers) obtained from **e** by fitting EDCs. Overlaid solid lines are guides to the eyes for the band dispersion. This exhibits the band deformation from parabola (red, 40 K) to W-shape (blue, 10 K) with decreasing the temperature.

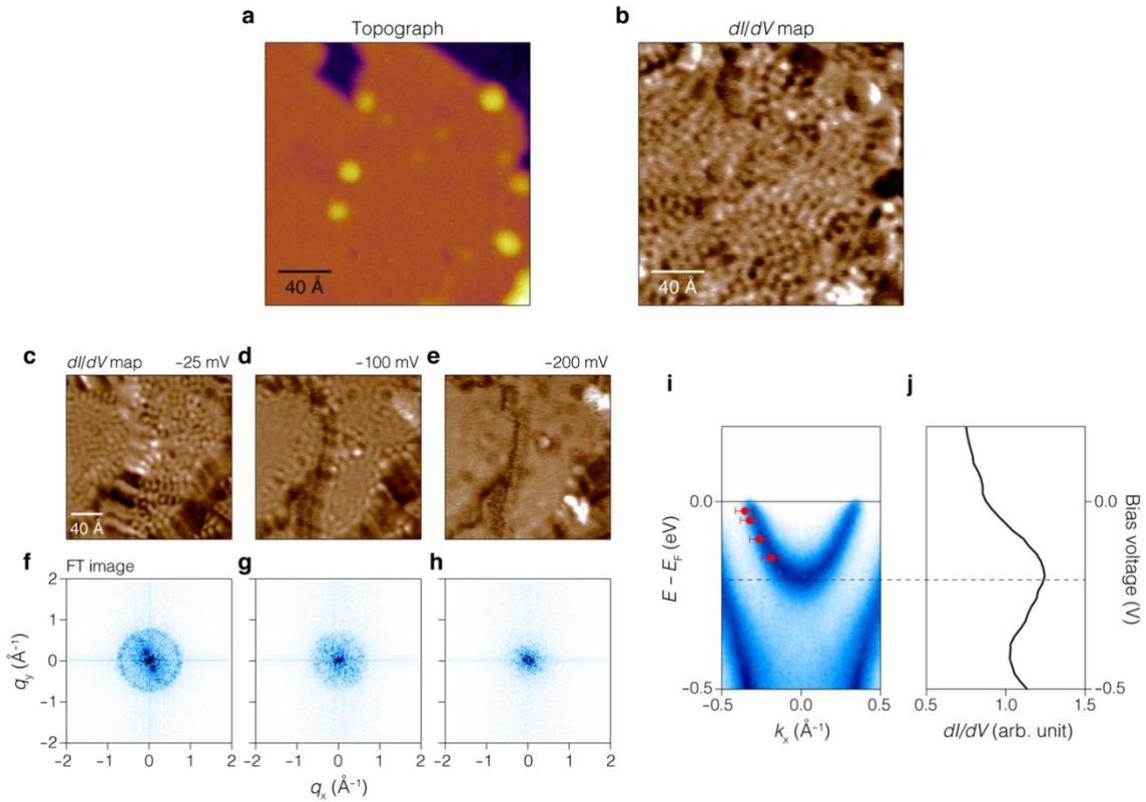

**Extended Data Figure 10 | Real space imaging of pure 2D electrons. a**, Topograph of cleaved $[Gd_2C]^{2+}\cdot 2e^-$ electride. The sample bias voltage ($V_B$) was set to $V_B$ = –25 mV. **b**, Differential conductance ($dI/dV$) map taken at same area with **a**. The periodic pattern shows standing wave of surface 2D electrons due to quasiparticle interference. **c–e**, $dI/dV$ maps observed at several different $V_B$ of (**c**) –25, (**d**) –100, and (**e**) –200 mV. The observed periodicity of standing wave increases with decrease of $V_B$. **f–h**, Corresponding Fourier transformed (FT) images of **c–e**. The scattering wavevector $q$ becomes smaller with decrease of $V_B$. **i**, Surface electron band observed by ARPES. Superimposed red markers are the half of the scattering wavevector ($q/2$) obtained from FT images at several different $V_B$ of –25, –50, –100, and –150 mV. **j**, $dI/dV$ spectrum of cleaved $[Gd_2C]^{2+}\cdot 2e^-$ electride. Dashed line shows that the energy of surface band minimum in **i** and a peak appeared in **j** match well. These results further confirm the pure 2D electrons formed on the top of $[Gd_2C]^{2+}\cdot 2e^-$ electride.